# PROTEIN FOLDING PROBLEM:
# SCIENTIFIC BASICS

Walter A. Simmons
Department of Physics and Astronomy
University of Hawaii at Manoa
Honolulu, HI  96822

The protein folding problem is stated and a list of properties that do not depend upon specific molecules is compiled and analyzed. The relationship of this analysis to future simulations is emphasized.  The choice of power and time as variables as opposed to energy and time is discussed.  A wave motion model is reviewed and related to the action in classical mechanics.  It is argued that the properties of the action support the idea that folding takes place in small steps.  It is explained how catastrophe theory has been employed in wave motion models and how it can be used in examination of successful simulations.

# INTRODUCTION

Proteins constitute the single most important group of molecules in the human body with some 20,000 proteins defined by function. The molecules are formed as linear strings of amino acids which promptly fold into biologically significant shapes. While the amino acid sequence can often be determined, deduction of the shape, especially in biological conditions, from the sequence is the essence of the protein folding problem.

It was discovered by Anfinson more than a half century ago that proteins of 350 residues or less can be unfolded and that they refold spontaneously to their native state when placed in water. That showed that the shape of the ground state and the mechanism to fold are stored in the amino acid sequence plus the water. That discovery also showed that folding is a problem in applied physics.

Because of the great importance of the shape and because of the large number of target molecules, simulations of folding are central to protein science. Recently, there have been some remarkable successes in folding simulation, especially using supercomputers. Simulations have also been successful in laboratories without supercomputers.

In this manuscript we report the compilation of a list of properties of folding molecules, (350 residues or less), which do not depend upon specific molecules or specific domains; we suggest calling these 'stylized facts' and classify them as 'basics' of folding. While such a compilation will not apply to every protein, the list of properties should be very useful in exploring successful simulations in order to probe aspects of folding that cannot be accessed experimentally.

After presenting our compilation and some related material, we analyze this collection of properties as a unit and point out some interesting interpretations.  Among our analytical comments we discuss our opinion that the power and time are an excellent choice of variables in addition to energy and time.  We then proceed to explain how these results either have been or could be used to interpret successful simulations.

Among the most striking features of protein folding is the fact that an enormous number and range of molecules fold to unique ground states in spite of small variations in solution conditions and in spite of the fact that most proteins are only marginally stable in the ground state.

In our earlier research, we argued that the stability must arise from some aspect of the underlying mechanics of wave motion on the molecule.  Using classical mechanics, (Lagrangian approach), we nominated the classical action for wave motion as the source of stability.  Following upon that, we used calculus of variations and catastrophe theory to seek a deeper physical understanding. In the last part of this manuscript, we update our analysis using the list or properties presented here. An important point is that the wave-motion-action approach is consistent with models in which folding takes place in small discrete steps.

It is perhaps worth remarking here that in some sense the stability of folding and the relative insensitivity to perturbations is the converse of the exquisite sensitivity of many protein interactions in biology.

Another topic discussed at least briefly include topology vs geometry; a subject of rapidly growing interest in theoretical physics, (we mention the new field of topological chemistry very briefly). We also remark that genetic research, particularly cancer research, has uncovered an incredible number of protein molecules that arrived without having the same screening through evolution that has shaped other molecules in our bodies; this redoubles the need for workable folding algorithms. We also mention briefly that artificial intelligence (AI) will become a powerful tool in folding.

Our most important conclusion is that the list of stylized facts and related theoretical ideas from various sources can be very useful in exploring the content of successful simulations.

# DEFINITION OF THE PROBLEM

The protein folding problem, as such, has been under intense research since about 1960. In this manuscript, we shall seek fundamental aspects of the folding process. We shall not address specific ground state structures in any detail.

Biological proteins of length less that about 350 residues spontaneously refold in solution to their unique native state (Wales). (We shall limit this manuscript to these relatively short proteins and leave out the subject of chaperones altogether.) The immediate objective is to obtain ground state structural information, which is of great value.

The sequence, together with the solution are assumed to contain all the necessary information to calculate the ground state structure. Protein folding is the premiere example of a self-organizing system, it probably engenders principles common to many lifeforms, and may shed light upon the origin of life and protein evolution. Therefore, in addition to structural information, an understanding of the mechanics of folding is also a priority.

To put the problem in clearer perspective, consider the complications that would arise if there were no limiting relationships or factors:

A.) Each residue has two dihedral angles. Therefore a sequence of three residues has six degrees of freedom between the ends. So, (aside from steric hindrance), the molecules can assume nearly any shape at all.

B.)   If we feed a length of smooth nylon line into a box, it will settle into a stable, tangled, condition; there would be many equivalent stable states.  Care would be required to put the line into a specific state such as a neat coil.  Protein molecules fold into just one final state.

C.)   The number of biological proteins is truly vast and the number following the simple pattern described above is astonishing.

D.)   Some of the earliest research, C1955, discovered that in solution, many sub-structures are near the threshold for thermodynamic instability.  No long, steep energy slope drives folding.  (Browning-motion driven models have been widely discussed.)

E.)   There are two dihedral angles in each which residue rotate independently of one another; their rotations commute.  That is not expected to be true of larger scale rotations and folds, especially where contacts form at specific points in the folding process.  This folding sequence dependence could result in many shapes for the ground state.  (While these dual folded structures are not unknown, they generally do not occur.)

F.)   Saddle points and bifurcations in the folding potential seem to present a problem.  The presence of any structures in the potential that divide the folding pathway will lead to multiple ground states, (which are not observed).

G.) The number of possible initial conformations is probably without limit and similarly the infinitely many possible trajectories. This large number of possible paths was the prime motivation for Cyrus Levinthal to argue that there must be pathways.

H.) Dissipation is a critical component of folding. Dissipation combined with almost random energy landscape would almost certainly lead to traps. However, traps are not generally found.

We add that in cases where dissipation does not use up all the energy by the time the molecule is near the ground state, the molecule anneals in the additional energy, which presumably could scramble structural information inherited during the early phase of folding.

I.) There are many forces involved in folding (Dill); some of the most important are non-local, (e.g. hydrophilic/hydrophobic forces), and generally speaking the energy available for folding is small. It is expected that each force changes with the details of the solution, (temperature, pH, etc.), yet folding is not easily modified by small changes in the solution.

J.) In the sequence space, there are large ranges of stability but also some points of extraordinary sensitivity to substitution mutations.

K.) The method of steepest descent works best if there is a single dynamical path. Branching paths give structures like lightning bolts; side branches burn out at random.

L.) Thermodynamics has no time-scale. Many aspects of thermodynamics are scale-invariant (e.g. temperature). This eliminates the possibility of a simple thermodynamic explanation for folding speed.

This list suggests that without some strong fundamental limitations in place, such as could be provided by a steep energy slope, there would be chaos. It seems extremely unlikely that some accident is involved for each and every sequence.

Search Space:

Another way to view the protein folding problem is to sketch out the magnitude of the search space which must be understood if any solution to the problem is to be regarded as satisfactory. Much has been written, for example from an evolutionary perspective (Koonin), (see below), about the combinatorics so we shall not repeat those points. We shall however point out some recent research on the number of different proteins, (by sequence), in the human body. In normal humans the number of functionally defined proteins is of the order of 20,000. There are individual variations so the number scales with the human population. If genetic related disease conditions are include the number increases.

(Lek) examined a sample of about sixty thousand individuals, focused upon the exome (protein coding region of the DNA), and found that one variant every in eight bases, on average.

In other research, (Hayward) studied 183 melanomas and found 20,894,255 substitution mutations.

It is clear from these figures and many others like them that a very highly efficient means of protein structuring is critically needed.

Sequence Code:

Much research has been devoted to aligning the various secondary structures with the sequence. It often happens that many different sequences align with the same structure.

Since many mutations do not change the function of a protein, mutations can compile into multiple sequences in different organisms.

In cases where the structure of a molecule is unknown but the sequence is similar to known proteins, then there has been great progress in calculating the unknown structure.

There are multiple, sophisticated, data bases that include structural data as well as related information such as experimental technique, organism, and, of course, references.

Protein Structures Distributions:

It would be appropriate for readers not familiar with protein structure to pause at this point and read the evolutionary paper by Koonin, Wolf and Karev, which we cited above. They explain the terminology (e.g. domains, families, etc.) and other relationships (orthologous, paralogous, etc.) and present numerous examples of distributions (e.g. power laws and Pareto distributions). We shall not cover that important material here.

Beyond Structuring:

The ground state, calculated out to the nearest sub-Angstrom, is the acid-test of folding models.  However, what is ultimately more important is the set of properties of the molecules in living systems.  The molecules have solubility, flexibility, they can diffuse through tissues and, most importantly they can react, usually with strong specificity, with other molecules.  While the ground state structure is usually a milestone in a calculation of molecular properties, that is not necessarily the most important result.  With a complete theory of folding, it may be practical to calculate many properties of the proteins in the biological environment without first obtaining the ground state structure.

The inverse of folding, i.e. unfolding, is possibly simpler than folding because the initial state is relatively fixed and usually known.  Because the molecules change in the cell, unfolding may be slightly more important than folding.

Broadly speaking, protein molecules have a strong skeleton involving covalent bounds, and secondary and higher structures involving weaker, (e.g. Hydrogen), bonds.  With the primary structure (sequence) fixed, the secondary and higher structures fold in solution.   Significant advances to our understanding has been made by applying quantum chemistry to certain structures.  Usually, but not always, these results are combined with methods of classical physics to simulate folding.

The protein folding problem breaks down naturally into two stages when one focuses upon a limited part of the molecule, (a single domain, perhaps).  One stage is the folding from denatured (open) form into a form near the native state.  The second is the final annealing, over ranges of Angstroms, to the final detail secondary and higher structures.  The first step can very likely be modeled using classical mechanics; the energy exchange with the solvent certainly destroys the coherence of quantum states at the temperatures and energies involved.  The second step very likely requires quantum mechanics and is likely to be exceptionally difficult to calculate; molecules are notoriously difficult to treat in quantum mechanics and these molecules have many degrees of freedom. (Chow)

The formation of contacts defines a further level of microscopic stages.  The formation of a contact will alter the shape of the chain in a crucial way.  Contact formation alters wave motion on the chain which has major consequences.  The Fourier representations, (i.e. spectral power densities), of the waves are altered due to the change in boundary conditions (please see further discussion, below).  Flow of power across the molecule is altered.  The mechanical linkages are altered implying a change in folding mechanism.

Setting aside the crystalline order and rigidity needed for x-ray diffraction determination of ground state structure, the states of the molecule slightly above ground might, optimistically, be considered as direct objectives of the folding problem.  To the extent that fluctuations and dynamic hydration can be measured in cells or in solution, those features present additional information sources for simulations.

The nuclear positions in the ground state are a priority but the electronic states, particularly surface electronic states, are probably important for extensions to explore biochemical reactions, molecular excitations and rearrangements. As far as is known, electronic surface states play only a minimal role in folding to the ground state structure but that is not certain. (As mentioned, above, the movement involving transverse waves usually have a minimum amplitude near the surface which could simplify some computations). Those surface states are almost certainly important in excited states and in chemical reactions.

We shall parenthetically introduce a minor caveat. The reaction coordinate interval between the cryogenic crystalline state of X-ray measurements and the biological state in room temperature solution with biochemical reactions taking place, has a phase transition. The crystal has a symmetry not present in the solution of molecules. The molecules in solution obviously have less rigidity and that freedom may be crucial to understanding protein function in solution and to understanding folding per se. Without a theory of folding, our ability to treat this phase transition is very limited. On the other hand, many experiments have explored this region and deduced features of folding dynamics.

## List: A list of some basic features of folding (stylized facts):

The following list draws together many current ideas in the literature of protein folding and includes some ideas and approaches from the literature of various other fields. The idea here, as discussed above, is to identify general features of folding which might allow us to identify some basic underlying features.

As originally emphasized by Levinthal, proteins fold very quickly from an unlimited number of open conformations, through an unlimited number of intermediate states to a unique end point. (Onuchic) A most remarkable phenomenon.

LIST:

Levinthal pathways (Levinthal).

Spontaneous and rapid folding ($N < 350$ residues) of biological proteins (Levinthal), (Mayor).

Unique ground state.

Insensitivity to solution perturbations.

The amino acid sequence codes ground state information.

Sequence plus water defines the folding mechanism.

Random sequences general do not fold to unique end-points (LaBean), (Bryngelson).

Nearly all proteins are constructed from a set of twenty amino acids.

Many common structures, (e.g. alpha helix, beta sheet, etc.) are found across protein molecules in the entire tree of life (Koonin).

Most mutations do not change the ground state structure or result in multiple ground states.

There are isolated high sensitivities to mutations (Shortle).

Proteins are marginally stable at room temperature (Dill 1990); usually, small energy available; some sub-structures marginally stable in solution.

Protein molecules are chiral: i.e. left-handed.

Various folding classifications have been identified; e.g. downhill folders, two-state folders, etc.

Principle of Minimal Frustration (Bryngelson).

Number of protein molecules is very large.

Failure to fold promptly to a unique state is associated with disease conditions; e.g. misfolding.

Evolution has imparted Pareto distributions to many statistical aspects of protein structure.

Additional notes:

Many excellent data bases are continuously maintained, e.g. PDB.  There is also much software for analysis available.

We remark that the existence of this list is, *per se*, a useful stylized fact; few items in the list are highly specific so the existence of the list points to there being general organizing principles in play. This list is no doubt incomplete and may have errors but it can be useful in structuring some folding theory research; particularly, as emphasized throughout these pages, in using successful simulations to explore possible mechanisms.

## Time, Energy, Power, & Simultaneity

A hypothetical chart of folding events with energy and time for each event along a pathway, would show events starting at various times and having various durations.  Some events would be simultaneous, some not so.  Timing and duration are probably critical to folding for some events and perhaps not for others.  Contact formation, for example, may require simultaneous release of energy into the backbone in multiple locations.  On the other hand, the annealing phase may be much less sensitive to simultaneous events because the structure has been partly set by that point; energy may determine the final structure without simultaneity.

Let us limit our thinking to simultaneous folding events taking place in a brief time window (short relative to overall folding time) and which are essential to folding.  The events take place across the molecule, or at least across a part of a domain.  In order that work 'scheduled' for this time window to be completed, there must be a minimum power available; that is there is a power threshold.

We can now begin to consider a putative Levinthal pathway as a sequence of time windows with corresponding power thresholds.

Many or all of the events may involve thresholds.  Note that energy thresholds and power thresholds are not the same in dissipative systems like folding.  A process may be allowed by energy but not allowed because the power is too low; (e.g. the energy is dissipated too soon). That problem can be offset by a flow of power from one part of the molecule to another.

Power flow makes it complicated to interpret the energetics of events in folding.

Simultaneity can be important for several fundamental reasons.  One is geometric and mechanical sequence control; e.g. arrangement and temporal sequence of contact formation.  Another reason is the addition of power (to promote natural pathway choice or suppress unnatural pathways).

Speaking more generally, it is a common experience that when a complex structure is assembled from small parts, the order of construction steps is critical.  We comment further on this point below.

Thresholds:

 An example of a global threshold can be appreciated using a simple analogy.   Suppose a simple computer operates on a time cycle and has a number of switches that toggle within a time cycle.  The minimum power that will always be sufficient is the power needed to toggle all of the switches during one time cycle.  With any less power, some computations will fail.  If the power is above the maximum, then some of it will not be used.

For folding events the time starts variously and has various durations. The maximum or threshold power is the amount needed to carry out all folds in the order required to form the ground state.

As we just said, the time for completion of some specific events, which do work on part of the structure, may be determined by what is happening elsewhere in the molecule. In those cases, the evens have important power thresholds.

Our main interest is in models that use wave motion to describe folding. Energy is the time integral of power and both quantities are useful in such models.

<u>Levithal Pathways:</u>

In the Threshold sub-section, we have introduced a putative pathway characterized by a power threshold.

The uniqueness of the pathway for a given initial conformation could arise due to the power being just the threshold amount at each time window; not more and not less. In analogy with the simple computer model a few pages back, sufficient power must be available to fold to a unique state in various solution conditions and various initial conformations.

By analogy, cloud to ground lightning bolts have bifurcations and horizontal strokes that are parallel to the ground and burn out before attaching to anything. This behavior is possible because of the excess power available along each stroke; multiple paths can be ionized simultaneously off of a single point in space.

The absence of folding paths that 'burn out' prematurely , (non-minimal frustration), and the absence of bifurcations leading to multiple ground states could be explained by the idea that the power available excites only the path with the lowest threshold.

Combining these two ideas about power suggests that there is a very narrow range.  Too much power and we get paths to meta-stable states off the folding trajectory and many end states that might be quasi-stable.  Too little power and some steps in the folding may not occur.

This idea of a narrow power range plays well with reproduction.

This power model, admittedly highly speculative, gives us our first indication of what a pathway might be; a flow of power along one of a set of equivalent trajectories.

## SIMULATIONS

Simulations, particularly those using especially engineered supercomputers, are having considerable success.  In view of the complexity of folding, simulations will always be a critical tool (Perez).  Hopefully, the ideas explored here will result in improvements to simulations.

Many of papers, listed as references, report work involving supercomputers and illustrate the range of successful projects recently reported.

Simulations performed in laboratories without access to supercomputers are very popular and it is worth remarking upon some ways those might discover some basics of folding. Once a successful simulation for a particular protein is in hand, that success can be exploited to look for features that might generalize.

An obvious approach is to look closely at changes to parameters that do not make any difference to a successful algorithm.  Having identified such a feature, one would, in general, look for a symmetry entailing some counter-vailing correction to the changes, calculational peculiarities that make the change irrelevant, or the presence of some dynamical stability mechanism.

In many cases, symmetry in physics is easy to explain.  A simple example may make that concept clear.  At a fundamental level, electricity has the (+,-) symmetry of charge. That symmetry pervades the subject; e.g. capacitors have two plates, the simplest antenna is a dipole, etc.

Note that while true symmetries are relatively easy to identify, broken symmetries, which may be just as important, can be very difficult to spot. In several branches of physics, there are successful theories that explain complex behavior using a simple parameter that somehow organizes the behavior. The parameter might be called an 'order parameter' and might have no specific external meaning or it might be called an 'adiabatic parameter' that facilitates understanding of the behavior through changes in structure that make no difference.

Another very useful analysis of simulations would be the statistics of contact formation. Both time and contact order are important.

How many Levinthal pathways are there? One for each initial configuration? More likely, the pathway is defined by a limited number of degrees of freedom, while other degrees of freedom fall into place without directly impacting the ground state shape, (perhaps by Brownian motion around a pathway).

Most simulations use some variation upon the method of steepest descent. The molecule is divided, computationally, into segments, the energies are calculated and the results combined. That is a sound approximation as far as it goes. Ultimately, we have to include long range forces and torques (i.e. vectors), which we discuss in another part, but limited-scale simulations could turn up some significant calculational artifacts.

Searching results of a successful simulation for possible 'on pathway' VS 'off pathway' processes could be carried out on most of the computers being used for folding; which is to say, seeking a definition of pathway.

Many researchers have drawn inspiration from the theory of phase transitions, which is mainly theory of symmetry. One aspect that can be explored in simulation is to notionally isolate a segment, say an alpha helix at a given position, and then to treat an ensemble of those with statistical physics (Zwanziger).

There is more information in a protein crystal than in one individual molecule. The process of near equilibrium crystal growth (or dissolution) might turn up length-scale correlations that tell us something about folding as seen through length-scale movements VS energy movement.  For example, phase changes are characterized mainly by symmetry and are often insensitive to small changes in material parameters.

Note that his sort of problem is more general than folding. Chemists who design solid structures encounter practical difficulties similar to what we have described above (Collins).  As in folding, solitons had a period of intense study but now are receiving declining attention; but they are not ruled out and successful folding solutions can be searched for evidence of solitons.

Correlations are closely related to noise analysis, which we mention in another Part.

Noise may be key, in some way, to folding (Huang). Simulations can manually introduce noise in various parameters and watch the impact upon the success of the algorithm. Either acute sensitivity to noise or utter insensitivity to noise would be particularly interesting.  It is also possible that noise stabilizes some otherwise unstable steps in folding, (Stochastic resonance).

Our viewpoint is that folding is a wave motion phenomenon, so tracking waves and wave packets in detail may turn up interesting processes.  In another part, we list some kinds of waves that may or may not be present in folding.  Searching successful simulations with this in mind might be very fruitful.

We have argued, above, that power transfer and simultaneity are critical concepts.  Even small computer simulations can map energy flow and look for convergences and simultaneous events.

As we have emphasized elsewhere, finding the ground state is the acid test of a simulation algorithm but detailed analysis of how and why a given simulation is successful on one molecule may be more important for non-supercomputer projects than extending the success to new molecules or new domains.

More Simple Strategies for Simulations:

With the availability of several very successful simulations of specific molecules, (and deeper explorations using supercomputers), here are several very simple, (and obvious), ideas that are difficult to explore by experiment but which can be tested using previously successful simulations (as discussed in the previous subsection).

Punctuation:

One of the simplest ideas from engineering is that carrying out a complex task in small stages, entails having an order for the steps and some sort of blueprint that includes signals for start and stop for each step.  In other words, there is punctuation (Gunasekaran).

Furthermore, in engineering, if some of the components are not suitably stable, then it makes eminent sense to do the steps one by one so that the work on one step does not destabilize a different step.  (I.e. black powder discipline).

Pursuing 'common sense engineering' further, we note that the greatest reliability is found in processes that have been used often; e.g. structures that appear across many molecules. The places where joints between proven sub-structures occur are the most likely to be sensitive to problems.

We don't know if evolution has employed any of these engineering ideas but it is reasonable to check. These ideas are not new but, as emphasized, simulations open interesting possibilities.

We divide the code into two parts: a structural part and a stop/start component. The latter code is probably very sparse compared to the first. The best evidence for that is that many mutations are accommodated without major structural changes. At the same time, conserved sequences point us to at least some information that cannot tolerate change.

As mentioned, the dramatic change in structure observed by Alexander, et al, (Alexander) may be the result of a mutation in a stop/start or event-sequence marker. In view of the dramatic nature of the change in structure, a marker for semi-stable process may have changed.

Since stop/start markers do not stand out, we note that dynamical variables, (energy, power, simultaneity), may be involved; as opposed to simple mechanical shifts. Moreover, we do not know if the punctuation marks are local or non-local. For example, some authors have proposed a twist VS writhe aspect to secondary structure (Bohr) that could obviously encode a punctuation mark. Recent hydrodynamic research (Moffatt) has shown that the sum of twist and writhe in flowing viscus fluids obeys a simple conservation law. That could encode a non-local signal.

Network scientists have carried these ideas much further (Gilarranz). They define modularity in terms of number of network connections local vs global and relate modularity to stability.

In the case of folding, the atomic level bond structure is not modified in folding but the shape is. In a later part we shall look into whether the 'steps' can be defined in terms of power flow.

# BACKGROUND TOPICS

The following short paragraphs are background.

Geometry VS Topology:

Scientific interest in topological concepts is currently growing explosively.  We summarize and briefly here.

The terms topology and geometry have very different meanings from each other in the physical sciences as well as in mathematics.  Geometry refers to anything that requires measurement of a dimensional quantity (length, energy, etc.).  Topology, in sharp contrast, deals with features that do not involve measurement.

We mentioned in the text that closing of a loop in the protein limits the oscillations that are possible and change the power spectrum.  The concept of 'loop', most especially the preservation of the 'loop' during various changes, is topological.  The details of the loop and changes that alter the 'loop' are mostly geometric.

A simple example along those lines can be constructed with a pen and paper.  Draw a smooth curve which forms a loop, crossing itself exactly once.  Now imagine that the loop is infinitesimally stretched and/or compressed in every direction within the paper; an arbitrary vector $(\Delta x, \Delta y)$ change at every point without breaking the curve.  Some aspects of the loop change: the length of the curve, the angles between the segments outside the loop, the area of the loop, its orientation, etc.  These are all geometric and are changes due to infinitesimal changes to dimensioned quantities. (The sum of the angles around the contact point, $2\pi$ , is also fixed but is geometrical (Weiner).

Some things do not change: the *existence* of the intersection, the *loop shape,* the *division* of the plane of the paper into two parts, (inside the loop and outside the loop): These do not depend upon geometry.

Such quantities are often said to be 'protected by topology'. In the above example, the two line segments that intersect each other, span a space of two dimensions – the same dimension as the paper. Topologically speaking, the lines cannot separate as they have nowhere to go.

Note that the length of the curve changes without changing the topological properties; this is an example of scale invariance.

Let us take this one step further. Suppose we replace the idealized loop with a real loop trapped in a two dimensional space, which is in contact with a thermal reservoir. Now the same topologically protected features are still there but they can carry energy and momentum. The point is that these features move about and carry energy but are, themselves, insensitive to the changes. For example, the loop still divides the plane into two parts; it cannot open up to facilitate motion not allowed by the confinement to two dimensions.

In this simple example, the loop topology is insensitive to thermal agitation.

Topology may be more important in folding than generally believed. Any topological characteristic cuts across all related dimensioned scales. Topology often manifests itself in stability against disturbances since such disturbances are dimensional. We shall draw the distinction between geometry and topology features frequently.

Suppose we have a trajectory of motion that is not changed by small external forces. Two extreme possibilities suggest themselves.  The structure is very rigid.  Or, there are many virtually indistinguishable trajectories present.  The first explanation is geometric, the second may be topological.

Some additional remarks beyond the scope of this Part

The topology of a system sometimes limits the possible energy and momentum structures.  Conversely, symmetries or constraints on energy may limit topology.

Scale invariance is usually associated with topology.  In the 'loop' example, above, changing the size of the loop (infinitesimally) makes no difference to the topological features; the distinction between inside and outside does not depend upon the size of the loop.

We remark that topology is involved in some of the most precise experimental determination of fundamental constants in physics; e.g. in the quantum hall effect.

As a closing remark for this box, we call attention to the very recently founded field called 'topological chemistry'.  In this case, energy directs the topology with interesting effects.

Electronics:

It is to be expected that before long new technology, such as free electron lasers, will make possible experiments on power and motion that are not now possible.  As has been emphasized above, simulations can attack a number of these problems immediately.

The linearity of many electronics concepts does not translate into folding.  However, the importance of critical points in dynamical variables may be of great importance in folding, even without the linear framework (e.g. differential equations and Laplace

Transforms). A critical point is one where one or more derivatives vanishes. We talk about this in some depth in Part II.

In electronics, a signal-processing device is usually analyzed in terms of two physical parts: the signal and the device. These parts are interrelated but can be treated separately to a useful degree. The signal, which in our case is the power, can be treated in idealized mathematical settings. The device, in our case the molecule, must be treated in its specifics.

Another observation from electronics is that analysis of power proceeds in two parallel lines: frequency domain and time domain. These domains can often be treated mathematically in an interchangeable way but in practice there are significant differences. For example, when the solution containing a denatured molecule is explosively diluted, there could well be a starting transient that is qualitatively different from the subsequent folding process. In electronics, starting transients often consist of short-lived, quiet, bursts of noise. We don't know anything about such noise in folding but it is clear that measuring it would provide us with useful information.

Artificial Intelligence:

Simulations can reveal natural patterns in the sequence in the form of rules. The hope of artificial intelligence is that the computer can find patterns and articulate them algorithmically. In those cases, the resulting algorithms might too complex for humans to appreciate but might nevertheless be very useful as algorithms, per se.

It is obvious that AI will play a major role in the future of folding simulations. In fact, it would come as no surprise if folding turns out to be one of the most productive areas of AI.The subject is outside the scope of these pages. We just remark that the AI technique of retro-propagating code to tune a model is probably already in use in folding. (Bohannon).

Informational Measures:

Shannon's definition of information is based upon resolution of ignorance. Suppose someone has flipped a penny and a nickel. A person who does not know the outcomes has this ignorance:

The probability of heads/tails for the penny is

$$P_P = 50\% / 50\%$$

For the nickel

$$P_N = 50\% / 50\%$$

and for the total

$$P_{Total} = P_P \bullet P_N$$

A message that resolves ignorance about the penny eliminates that uncertainty and similarly for the nickel. Therefore, the information depends, in some fashion, upon the probabilities that were removed,

$$I_P = I_P(P_P)$$
$$I_N = I_N(P_N)$$

If the information is defined to be additive then

$$I_{Total}(P_P \bullet P_N) = I_P(P_P) + I_N(P_N)$$

The choice of solution to this equation, for the dependence of information upon probability, is

$$I(P) = -\frac{1}{2}Log_2(P)$$

Using the number of possible amino acids (twenty) and the number of residues in a particular molecule lets us calculate the information content of the sequence (in isolation).

Research into 'looking backward' from folding of contemporary proteins toward the origin of life and evolution is still an open field. Since the uniqueness of folding is intimately related to reproduction in contemporary biology, we cannot exclude the possibility of undiscovered signals from the distant past encoded in the information contained in the sequence. When we have a full understanding of the sequence code for folding, whatever is left can be investigated with interest.

If most of the information in the sequence code is related to managing the folding in detail and the start/stop code is completely independent and very sparse, one would expect that plots of structure VS anything directly related to domains would show two features. We know of no such features, see (Koonin). That may mean that there is no start/stop code or it is very similar to the structural code.

This is a research topic in which the methodology of information science may be applicable.

# RECAPITULATION OF PART I

The first part has been mostly descriptive.  We are seeking properties of protein folding that do not depend upon just a limited number of specific molecules. Facts and ideas were drawn together from the literature of folding and other subjects and presented from a different perspective.

The primary motivation is the striking stability of folding; a very large number of molecules with varying sequences follow the same pattern of folding to unique ground states in spite of various solution conditions and in spite of the fact that relatively little energy is available to drive the folding.

Noting that successful computer simulations of specific molecules are now appearing regularly, we emphasized general physical aspects of folding that might be inaccessible to present day experiments but amenable to exploration with simulations – especially using popular simulations from laboratories without access to super-computers.

We inserted miscellaneous background ideas into this text. These included borrowed ideas from engineering, information science, and other fields.

We note that power flows have several interesting properties that can be studied in simulation.  For example, contrast folding power flow to the power flow in lightning striking between clouds and ground.  We observe that lightning strokes frequently bifurcate. At the point of bifurcation there is sufficient power to drive a current flow down two distinct paths simultaneously.  That is not the case in folding; side paths mean misfolding and they are not common.

If we imagine certain junctures where there is sufficient energy in the molecule to trigger either or both of a pair of alternative folding steps, but where the power is sufficient for one but not for the other, then we have a possible explanation of the uniqueness of the pathway. (Of course, parallel and equivalent paths almost certainly occur.)

Our perspective in this manuscript and our earlier research is centered upon wave motion. A putative pathway might have a bandwidth limitation and/or a resonant frequency. In these cases, a given pulse of energy may not transfer enough power to drive the molecule down the putative pathway but rather the pulse might move along to do work elsewhere.

Conversely, going into a particular folding step insufficient energy may be inherited from the prior phases but fresh energy might be released promptly to provide sufficient power for the step. Potentially, the molecule may be pushed past a potential energy trap more quickly than it can descend and lose energy by dissipation. Experiments to measure these simple physical processes are difficult, but simulations are easy.

When any complex structure is constructed from various smaller components it is likely that the assembly process must take place in relatively well defined stages. Moreover, there is a question of whether a given step can be initiated before some set of previous steps is complete. An additional issue here is that sub-structures in some cases are not particularly stable; (in fact, many helical structures are thermodynamically unstable in solution). How is it possible for a complex structure to self-organize if the components are not strongly stable?

The wave picture that we will elaborate upon in Part II offers possible answers.

The simplest answer to the questions just posed is that various stages go to completion before succeeding stages initiate. E.g. available power might be focused upon bring unstable parts together. It is unlikely that this is a universal answer but it can be explored in simulation.

Based upon these simple ideas for model construction and upon the success of Levinthal pathways we next suggested that the pathways are defined by power flow through the multi-dimensional configuration space.

We emphasized in this part, what is already well known, that there is a great deal of important information in the spectra of waves on the molecule. Simulations to understand the role of wave power spectra will be very valuable. For example, Browning motion, with a spectrum $1/f^2$ , favors low frequency, $f$ , and will usually move side chains in non-specific directions. A wave pulse with a wide spectrum may drive motion in a more specific direction.

Part II, below, takes up classical mechanics fundamentals.

## Part II

## Introduction

From part I, we inherit the conjecture that the Levinthal pathways are collections of equivalent trajectories which may describe conduits for power flow.  We make use of that idea here.

While part I was mostly driven by empirical information and simple conjectures, this part approaches folding rather differently.  Part I was a survey and collection of stylized facts about folding, generically.  We worked toward the goal of identifying some common features of many protein molecules and organizing those facts using relatively simple ideas from other branches of science and engineering.  In this part, the goal is to propose possible origins of these common features in classical mechanics; in particular in the action, which has been central to our own research program for several years.

This part enters the fields of symmetry, topology, stationary action, calculus of variations, higher stabilities of the action, and aspects of catastrophe theory.  We shall not re-derive any material in the literature, instead we describe these to inform interested readers, especially students, of these possibilities.

## FRAMEWORK FOR OUR ANALYSIS:

We began with a definition of the problem and with a detailing of many interesting aspects of folding. Most of the descriptive (experimental) material here is in the form of stylized facts. That is, we make statements that guide thinking but which probably have exceptions. For example, we often mention that each of these (small) biological protein molecules folds to a unique ground state; in fact, not all biological molecules do so; a few do not.

A central issue is the existence and properties of the Levinthal pathways. We assume for this manuscript that pathways exist. Many models of various kinds explicitly or implicitly assume the existence of some 'pathways', though not all models are making the same assumptions. Our assumption is that the folding follows some, yet to be elucidated, pathway from an open denatured state to a unique ground state.

We make the assumption that amongst the various motions of the folding molecule, the most important are torsional waves together with weak chemical reactions at the contacts.

While we display few specific calculations in this manuscript, we refer often to calculations. Except as mentioned, we use only classical mechanics. Our general assumption is that because of our assumption of wave motion, Lagrangian mechanics is used. The principle reason for that choice is that it directly relates a starting condition to an end state. Also, it is particularly useful where symmetries apply. That method has been extended to include some aspects of dissipation.

Perturbation Theory:

Modeling motion by choosing appropriate increments, (e.g. steepest descent), recalculating, and repeating the procedure is a form of perturbation theory. It has been commonly used in simulations with some solid successes but whether it works or not in general is one open question about fundamentals.

In particular, near a saddle point, it may be computationally difficult to select the 'right' increment. It would not be a surprise if saddle points introduce acute sensitivity to initial conditions. In the case of non-local forces, especially wherein we do not know the range of such forces, it is conceptually difficult to choose the 'right' increment.

The opposite of ambiguity, as with saddle points and non-local forces, is what is called 'transversality', which will be described below. In that case strong restrictions virtually eliminate choices.

We recall that, as we said before, the dynamic must encompass discontinuous jumps in the sequence space.

Classical Mechanics and Classical Statistical Mechanics:

Besides the direct application of $\vec{F} = m\vec{a}$ , there are two commonly used approaches to classical mechanics calculations; Hamiltonian and Lagrangian. These approaches are equivalent, but in some cases there are conceptual advantages of one over the other two. The Hamiltonian approach is often most useful when one wants to calculate the detailed time development of a physical system, it is less useful in dissipative systems except under certain circumstances.  The Lagrangian approach is particularly useful in cases where there are constraints, where one is primarily interested in the relationships between initial and final states, and in cases where there are symmetries.  The approach has been extended to include dissipative processes.

We shall use the Lagrangian approach exclusively.  The geometry of the molecules obviously places constraints on all variables involved and the Levinthal pathway idea is, itself, a statement implying constraints of a dynamical sort.  The concept of Levinthal pathways, (assuming several pathways in a given molecule), certainly suggests some sort of symmetry in the dynamics.

Generalized Coordinates:

This concept, from Lagrangian mechanics, refers to coordinates that take into account various constraints.  More specifically, it is reasonably to suppose that Levinthal pathways somehow define some coordinates that are favored; i.e. they correspond to generalized coordinates.

A reminder: if a bead moves without friction along a curved horizontal wire, distance along the wire corresponds to a generalized coordinate.  There are forces perpendicular to the wire but they do no work and may be disregarded.

Waves:

Much research has gone into wave motion during folding. Michael Levitt has published some beautiful simulations of molecules ringing at wavelengths up to the same size as the molecule.  Wave packets may be more important than specific Fourier components.  Wave packets are localized and spend a finite time in any given location; the power transferred between a wave packet and a part of the molecule is undoubtedly of fundamental importance.  It is very likely that all wave propagation in molecules is highly dispersive, resulting is rapid distribution of power.

Moreover, waves of long wavelength accommodate to, and therefor communicate, structural information over various distances.  For instance, most residues are chiral and that would certainly influence the structure of mechanical waves.  Glycine is not chiral but a molecule with few glycine residues could still have the chiral structure as the majority residues.

Waves on a protein can be torsional, compressive, or transverse; (we do not consider shock waves here).  Waves can also propagate through the solution. The boundary conditions imposed upon the waves change during folding.  We shall consider that folding takes place by means of waves, particularly torsion waves.

The mechanical movement of the backbone results in collisions of the chain with itself.  The underlying Fourier spectrum is expected to change when these collision events occur; this seems to define sub-phases of folding.

The formation of a contact during, or due to, the passage of a wave packet will alter the spectrum of the packet and transfer long wavelength energy out of the wave.

We remark that torsional waves can transmit torque over long stretches of the molecule contributing to non-locality.

Wave motion, as just mentioned, entails traveling waves, standing waves, interference, dissipation, etc.  All of these are conveniently described by the action.  Furthermore, there are many technical methods for dealing with wave motion; Fourier and Laplace transforms, Fluctuation-Dissipation theorem, Onsager relations, etc., as well as theoretical principles to be applied; rotation invariance, time-reversal invariance, reflection non-invariance, etc.

There are multiple scales upon which waves travel in folding. The most important break point between scales is the length of a residue. Mechanical waves on a longer scale can propagate generally but for wavelengths shorter than the scale of the residues such mechanical waves are limited; however, such wave motion can occur in the solution and couple distant residues.

The most important pieces of information that can be readily communicated by wave motion are length and time; wavelength and period. We note that in a different field, (DNA nanotechnology), organized communication by structure change propagation has been demonstrated.

In the spirit of a question posed in an earlier part, should we ask if all the different wave forms, modified as they are by chain collisions and changing boundary conditions, somehow conspire to result in smooth folding to a unique state, little perturbed by various changes in solution conditions?

Chirping:

Mechanical waves on the molecule are generally limited in wavelength by the overall dimension of the structure at any given time. Since the structure is collapsing during folding it follows that the power in the waves will migrate to shorter wavelengths. In unfolding, the reverse may take place.

This may be related to a variety of features that are specific to protein reaction dynamics. (Karplus (2000))

A similar, but more specific phenomenon in any kind of electromagnetic radiation, is called 'chirping'. It usually involves a change in dispersion as well as frequency. Chirped electromagnetic waves have been used in analytic instruments for biomolecule studies. (Kubitzki), (Ariese)

The migration to shorter wavelengths suggests that more power becomes available for changing smaller scale structures. The amount of energy stored in the molecule at any time could be characterized by a 'Q' value, after the Johnson Q value in physics. That is a measure of the bandwidth and the losses in a vibrating system. In simple systems the Q provides a guide to limitations on power; applying more power at some point does not result in more stored energy.

We are unaware of any searches for mechanical wave chirping or Q-value analysis in protein-folding simulations.

Action:

The concept of action is normally described in the first few pages of any classical mechanics textbook. It will be defined below and used extensively in later papers, to quantify folding dynamics. It is to be anticipated that the action has discontinuities when collisions take place and/or contacts form.

We have emphasized in the manuscript, from the title page onward that we seek to understand folding at its most basic level. That level is the action that describes the underlying mechanical motion, which we assume is mostly torsion wave motion.

Part II deals with the theory of action in folding.

## Dynamical Fundamentals Theory

The exposition, in Part I, of putatively common features of the folding of a wide range of molecules may be tied together in a pattern of physical properties. The way that we address that is to refer to the action of classical mechanics, always assuming an underlying torsional wave mechanism. For simplicity, we shall focus upon one particular sub-set of stylized facts: the folding is rapid, relatively stable against changes to solution conditions, and leads to a unique ground state.

We shall assume, as is common, that we can assume the existence of a potential per unit volume function that incorporates the molecule and the solution. We shall call this $V$.

There are two classes of arguments of the potential. The first is dynamical variables, (defined as generalized coordinates), such as bond angles and distances between atoms. The second is specific parameters such as, say, the molecular weight and moment of inertia of a specific side chain.

The generalized coordinates are not known and we do not address here the complicated problem of including all the possible constraints.

From the potential and the corresponding kinetic energy, $T$, we construct the Lagrangian density, $L = T - V$.

From that, we construct the action, (symbolically),

$$S = \int L(\overline{x}) dx dt$$

The integration over space and time is taken from the initial to final states. The dynamical variables are symbolized by $\overline{x}$. The variation of possible solutions is symbolized,

$$\delta S = 0$$

which says that any small change to the dynamical variables leaves the action invariant. The Euler-Lagrange differential equations for the motion follow directly.

Evolution has introduced interrelationships between the additional parameters and we observe the results as stylized facts. Our task here is to see how adjustments to the parameter set might have introduced the high degree of stability that we see in folding.

The folding of proteins to unique final states is obviously related to reproduction. We shall conjecture the a primary target of evolutional development has been folding to a unique state, independently of minor changes to the solution conditions.

The first question is, what parameters can evolution have modified? The simplest answer is that evolution chose specific sets of amino acids from amongst all those amino acids available in nature, and, within those sets, chose specific sequences.

To understand how to solve this problem, we turned to calculus of variations. One standard approach is to vary our parameters in a second or third variation of the action.

Deeper stability is defined as

$$\delta^2 S = 0$$

This implies the existence of corresponding partial derivatives of the potential, a matrix in which some elements have

$$V^{''} = 0$$

For both, or either, diagonal of off diagonal elements.

To further understand the relationship between these critical points of the potential and the resulting stability, we turn to a powerful theory of stability, catastrophe theory.

Again, we are assuming wave motion. From, (Gilmore) we find that the convergence of a wave field is actually governed by the critical points of the corresponding action. More specifically, if the action has vanishing first and second derivatives for some coordinates, then the solution the differential equations will converge very sharply for coordinate values that satisfy the above conditions.

Formation of contacts is poorly understood. We conjecture that the action describes the motion between contacts. Dill has observed (Dill) that the Levinthal problem can be understood if the folding occurs in steps of a suitable nature. There may be punctuation demarking the steps or the contacts might mark the steps. In our action based picture, the process takes place in steps between punctuation marks which might be contacts.

This Part is a very high level view of our ongoing research. More details are to be found in the literature (Simmons) et sequence.

## CONCLUSIONS

We began by collecting together stylized facts about protein folding that seem to apply rather generally; (as opposed to aspects of specific proteins). We reported some conclusions from our earlier research and we particularly emphasized that the availability of successful simulations of some molecules presents an opportunity to explore aspects of folding that are currently inaccessible via experiment.

Important among these are power and simultaneity.

A short list of these stylized facts are:

- From the geometry of protein chains, we note that the molecule can bend into almost any shape.

- Folding in water leads to a stable and unique ground state structure in spite of variations in solution conditions.

- Some sub-structures are near the threshold of thermodynamic instability in solution.

- Almost all proteins are constructed from a set of twenty amino acids.

- Most mutations do not disrupt folding but some do.

Our approach has been to impute general behavior of folding proteins to aspects of the action (classical mechanics) which can be explored either analytically or via simulation. We then explored how these common features would be manifest in the action.

Starting with stability and a torsional wave model of folding, we argue that the motion is controlled by critical points in the action, as described in calculus of variations and catastrophe theory. This picture is consistent with the relatively weak stability of sub-structures in solution, the common stability of structure to mutation, the contrasting, but occasional, exquisite sensitivity of structure to mutation.

Simulations should be studied carefully for combinations of parameters that do not disrupt the ground state structure when perturbed.  Such combinations of parameters signal the stability found in our mathematical analysis.  Our action model tells us that these combinations of parameters are associated with critical points in the potential.

Wave motion will change when each contact forms because a new contact changes the boundary conditions on the waves.

In concert with the theory of Dill that folding occurs in steps, we suppose that some sort of punctuation marks are present; these may have some connection to contacts.  We suggest that mutations do not strongly impact structure during steps but may have violent effects upon punctuation marks.

For an interesting discussion of structural relationships, see (Englander).

These summary statements about steps VS stability of the ground state and the presence of sub-structures with marginal stability in solution, as well as numerous other statements in our text, can be explored in simulations.

## Acknowledgement

J.L. Weiner, Department of Mathematics, University of Hawaii at Manoa is co-author of several papers cited.  The author also thanks Professor Weiner for reviewing some parts of this manuscript.

## Notice

Comments are invited to the author by email at:

WSIMM2  AT  Phys.Hawaii.edu

BOOKS, REVIEWS, and PAPERS

# GENERAL REFERENCES

Effects of temperature on protein structure and dynamics: …
Biochemistry (1992), 31, 2469

(Zwanzig) R. Zwanzig, "Two-state models of protein folding
kinetics", PNAS <u>94</u>, 148 (1997).